# Realizing optical pulling force using chirality

Kun Ding,[1,2] Jack Ng,[1,3,*] Lei Zhou,[2] and C. T. Chan[1,†]

[1]*Department of Physics and Institute for Advanced Study, The Hong Kong University of Science and Technology, Clear Water Bay, Hong Kong*

[2]*Department of Physics, Fudan University, Shanghai, China*

[3]*Department of Physics and Institute of Computational and Theoretical Studies, Hong Kong Baptist University, Kowloon Tong, Kowloon, Hong Kong*

**Abstract**

We derived an analytical formula for the optical force acting on a small anisotropic chiral particle. The behavior of chiral particles is qualitatively different from achiral particles due to new chirality dependent terms which couple mechanical linear momentum and optical spin angular momentum. Such coupling induced by chirality can serve as a new mechanism to achieve optical pulling force. Our analytical predictions are verified by numerical simulations.

A great majority of research in optical micro-manipulation focuses on the interaction of light with isotropic spherical particles, such as polystyrene, glass, or metallic beads [1,2,3,4,5,6]. These simple spherical particles serve as prototypes for a simple intuitive understanding of light-matter interaction [7,8,9]. However, particles in nature typically have lower symmetries and the broken symmetry can give rise to new physics [10,11,12,13,14,15,16,[17]18,19]. In practical applications, we may need to deal with aggregates of particles with complicated shapes and geometry [20,21,22]. The underlining physics for these more complicated particles or their aggregates are not well understood, although it is generally assumed that the physics is not that different from a spherical entity. In this paper, we consider the optical micro-manipulation of a class of chiral particles. Unlike simple spherical beads, the absence of reflection symmetries in chiral particles leads to a series of fascinating phenomena and in particular, the chirality allows light to attract a chiral particle.

We shall first analytically derive the optical force expression for a small anisotropic chiral particle in general. Then, in order to capture the physics without unnecessary complications, we consider the special case of a small isotropic chiral particle, where the analytic description is much simpler and easier to interpret. We shall then see that new physics (chirality dependent terms that couple mechanical linear momentum and angular momentum of photons) and new implications (optical pulling force) emerge owing to the chirality of the particles. The possibility of using chirality to achieve optical pulling forces is then verified by accurate numerical simulations based on scattering theory.

A chiral particle is one whose mirror image cannot overlay on the original particle and the chirality of a particle is manifested when it interacts with another chiral entity. Electromagnetic waves can also have chiral character, and for a monochromatic wave, its chirality can be defined as [23,24,25]

$$K_{Chirality} = \frac{\varepsilon_0 \omega}{2} \text{Im}\{\mathbf{E} \cdot \mathbf{B}^*\}, \qquad (1)$$

while the time averaged chirality flux is given by

$$\langle \mathbf{\Psi} \rangle = \langle \mathbf{\Psi}_e \rangle + \langle \mathbf{\Psi}_m \rangle, \qquad (2)$$

where

$$\langle \mathbf{\Psi}_e \rangle = \frac{\mu_0 \omega}{4} \text{Im}\{\varepsilon_0 \mathbf{E}^* \times \mathbf{E}\},$$
$$\langle \mathbf{\Psi}_m \rangle = \frac{\mu_0 \omega}{4} \text{Im}\{\mu_0 \mathbf{H}^* \times \mathbf{H}\}, \qquad (3)$$

are the part of the chirality flux expressed in terms of the electric and magnetic field, respectively. We note that $\langle \mathbf{\Psi}_e \rangle$ and $\langle \mathbf{\Psi}_m \rangle$ are proportional to the time averaged spin density in the paraxial limit [26].

The left and right circularly polarized plane waves are of opposite chirality, which will interact with a chiral particle in a different way. This can lead to different or even opposite optical forces. It is important to note that additional chirality dependent terms must be added to the typical expressions for optical force acting on achiral particles. Our task is to highlight the unique role of these chirality dependent terms in optical manipulation.

To illustrate the essence of the physics, it suffices to consider a small chiral particle characterized by an induced electric dipole moment $\mathbf{p}$ and a magnetic dipole moment $\mathbf{m}$, which are given by [27]

$$\begin{aligned}\mathbf{p} &= \vec{\boldsymbol{\alpha}}^{ee}\mathbf{E} + \vec{\boldsymbol{\alpha}}^{em}\mathbf{B} \\ \mathbf{m} &= -\vec{\boldsymbol{\alpha}}^{emT}\mathbf{E} + \vec{\boldsymbol{\alpha}}^{mm}\mathbf{B}\end{aligned}, \quad (4)$$

where the superscript $T$ denotes transpose, $\vec{\boldsymbol{\alpha}}^{ee}$, $\vec{\boldsymbol{\alpha}}^{mm}$, and $\vec{\boldsymbol{\alpha}}^{em}$ are respectively the electric, magnetic, and chiral polarization tensors of the particle. A chiral particle is characterized by having a non-zero $\vec{\boldsymbol{\alpha}}^{em}$, which can be due to the chirality of the constituting molecules or the chiral shape of the particle. While the chirality of naturally occurring substances is usually weak, artificial nano or micro-structures can possess gigantic chirality [28,29,30,31]. Substituting (4) into the time averaged optical force expression in Ref. 32, one arrives at (see supplementarl material)

$$\langle \mathbf{F} \rangle = -\nabla \langle U_f \rangle + \frac{1}{4}\mathrm{Re}\left\{ \begin{array}{l} \mathbf{T}^{ee}:\vec{\boldsymbol{\alpha}}^{ee} + \mathbf{T}^{em}:\vec{\boldsymbol{\alpha}}^{em} \\ +\mathbf{T}^{mm}:\vec{\boldsymbol{\alpha}}^{mm} + \mathbf{T}^{me}:\vec{\boldsymbol{\alpha}}^{em,\mathrm{T}} \end{array} \right\} - \frac{k^4}{12\pi\epsilon_0 c}\mathrm{Re}\{\tilde{\mathbf{L}}+\tilde{\mathbf{S}}\}, \quad (5)$$

where $\langle f \rangle$ denotes the time averaged value of $f$,

$$\langle U_f \rangle = -\mathrm{Re}(\alpha_{ee})\frac{\langle u_e \rangle}{\varepsilon_0} - \mu_0 \mathrm{Re}(\alpha_{mm})\langle u_m \rangle - \mathrm{Im}(\alpha_{em})\frac{K_{Chirality}}{\varepsilon_0 \omega} \quad (6)$$

is a scalar which can be interpreted as the time averaged free energy, and

$\langle u_e \rangle = \varepsilon_0 |\mathbf{E}|^2 / 4$ and $\langle u_m \rangle = \mu_0 |\mathbf{H}|^2 / 4$ are the electric and magnetic energy density.

The tensors $\mathbf{T}^{ee}$, $\mathbf{T}^{em}$, $\mathbf{T}^{me}$, and $\mathbf{T}^{mm}$ have components of the following form

$$T^{ee}_{n,ji} = \left[ (\partial_n |E_i|)|E_j| - (\partial_n |E_j|)|E_i| - i|E_i||E_j|\partial_n(\Theta_i + \Theta_j) \right] e^{i(\Theta_j - \Theta_i)},$$
$$T^{em}_{n,ki} = \left[ (\partial_n |E_i|)|B_k| - (\partial_n |B_k|)|E_i| - i|E_i||B_k|\partial_n(\Theta_i + \Phi_k) \right] e^{i(\Phi_k - \Theta_i)}, \quad (7)$$

$$T^{mm}_{n,ji} = \left[ (\partial_n |B_i|)|B_j| - (\partial_n |B_j|)|B_i| - i|B_i||B_j|\partial_n(\Phi_i + \Phi_j) \right] e^{i(\Phi_j - \Phi_i)},$$
$$T^{me}_{n,ki} = \left[ (\partial_n |E_k|)|B_i| - (\partial_n |B_i|)|E_k| + i|B_i||E_k|\partial_n(\Phi_i + \Theta_k) \right] e^{i(\Theta_k - \Phi_i)}, \quad (8)$$

where $\Theta_j$ and $\Phi_j$ are the phases of the electric and magnetic fields, respectively, and

$$\tilde{\mathbf{L}} = \mathbf{E} \cdot \left( \vec{\boldsymbol{\alpha}}^{em*} \times \vec{\boldsymbol{\alpha}}^{ee} \right)^T \cdot \mathbf{E}^* - \mathbf{B} \cdot \left( \vec{\boldsymbol{\alpha}}^{mm?} \times \vec{\boldsymbol{\alpha}}^{em} \right)^T \cdot \mathbf{B},$$
$$\tilde{\mathbf{S}} = \mathbf{B} \cdot \left( \vec{\boldsymbol{\alpha}}^{em*} \times \vec{\boldsymbol{\alpha}}^{em} \right)^T \cdot \mathbf{E}^* - \mathbf{E} \cdot \left( \vec{\boldsymbol{\alpha}}^{em?} \times \vec{\boldsymbol{\alpha}}^{ee} \right)^T \cdot \mathbf{B}. \quad (9)$$

It is clear that all terms that depend on $\alpha_{em}$ are related to chirality.

While Eq. (5) is a general expression for a small chiral particle, to grasp the physics, we consider an idealized particles with isotropic polarizabilities, i.e. $(\vec{\boldsymbol{\alpha}}^{ee})_{ij} = \alpha_{ee}\delta_{ij}$, $(\vec{\boldsymbol{\alpha}}^{em})_{ij} = \alpha_{em}\delta_{ij}$, and $(\vec{\boldsymbol{\alpha}}^{mm})_{ij} = \alpha_{mm}\delta_{ij}$. Equation (5) is simplified to (see supplemental material)

$$\langle \mathbf{F} \rangle = -\nabla \langle U_f \rangle + (C_{ext} + C_{recoil}) c^{-1} \langle \mathbf{S} \rangle$$
$$+ \nabla \times \left[ -(C^{\mathbf{p}}_{ext} c / \mu_0 \omega) \langle \Psi_e \rangle - (C^{\mathbf{m}}_{ext} c / \mu_0 \omega) \langle \Psi_m \rangle + \mu_0 (\mathrm{Re}\,\alpha_{em}) \langle \mathbf{S} \rangle \right]$$
$$- \frac{1}{\mu_0 \omega} \left[ 2\omega^2 \mu_0 (\mathrm{Re}\,\alpha_{em}) - \frac{k^5}{3\pi\epsilon_0} \frac{\mathrm{Im}(\alpha_{ee}\alpha^*_{em})}{\epsilon_0} \right] \langle \Psi_e \rangle \quad (10)$$
$$- \frac{1}{\mu_0 \omega} \left[ 2\omega^2 \mu_0 (\mathrm{Re}\,\alpha_{em}) - \frac{k^5 \mu_0}{3\pi\epsilon_0} \mathrm{Im}(\alpha_{mm}\alpha^*_{em}) \right] \langle \Psi_m \rangle$$

where

$$C_{ext} = C^{\mathbf{p}}_{ext} + C^{\mathbf{m}}_{ext} \quad (11)$$

is the extinction cross section, and

$$C_{ext}^{p} = \frac{k\text{Im}(\alpha_{ee})}{\epsilon_0}, \quad C_{ext}^{m} = k\mu_0 \text{Im}(\alpha_{mm}) \tag{12}$$

are, respectively, the part of extinction coefficient caused by the electric and magnetic dipoles, and

$$C_{recoil} = -\frac{k^4 \mu_0 \text{Re}(\alpha_{ee}\alpha_{mm}^*)}{6\pi\epsilon_0} - \frac{k^4 \mu_0}{6\pi\epsilon_0}|\alpha_{em}|^2 \tag{13}$$

is directly related to the recoil force. Note that in (10), we have threw away the small terms that are proportional to $\text{Re}\{\alpha_{em}\}$ or $(\text{Re}\{\alpha_{ee}\}\text{Im}\{\alpha_{mm}\} - \text{Im}\{\alpha_{ee}\}\text{Re}\{\alpha_{mm}\})$.

To make things even simpler, we consider an incident paraxial beam. In that case,

$$\langle \mathbf{L}_s^m \rangle = \frac{\mu_0}{4\omega i}\mathbf{H} \times \mathbf{H}^* = \frac{\epsilon_0}{4\omega i}\hat{\mathbf{k}}[\hat{\mathbf{k}} \cdot (\mathbf{E} \times \mathbf{E}^*)] = \langle \mathbf{L}_S^p \rangle \equiv \langle \mathbf{L}_s \rangle, \tag{14}$$

which leads to a compact formula for optical force:

$$\langle \mathbf{F} \rangle = -\nabla \langle U_f \rangle + (C_{ext} + C_{recoil})c^{-1}\langle \mathbf{S} \rangle + C_{ext} c\nabla \times \langle \mathbf{L}_s \rangle$$
$$- \frac{k^5}{3\pi\epsilon_0^2} \text{Im}[\alpha_{em}^*(\alpha_{ee} + \frac{\alpha_{mm}}{c^2})]\langle \mathbf{L}_s \rangle, \tag{15}$$

We note that the first and last terms of Eq. (15) are both chirality dependent, and do not exist in the expression for a non-chiral particle [33,34]. For the same electromagnetic field characterized by the same $K_{Chirality}$, the force of the particle depends linearly on $\text{Im}(\alpha_{em})$, which has opposite sign for particle with opposite chirality.

Next, we give an example which manifests the novelty of the chirality dependent force terms. Recently, there are increasing interest in the so called "optical tractor beam," where a forward propagating beam can drag a particle backward [35.36,37,38,39,40,41]. Examples of optical tractor beams include the optical solenoid beam [40,41] and the optical pulling force [35,36,37,38,39,40,41]. The former has its backward force originates from the gradient force [40], whereas the backward force of the optical pulling force is a *negative* scattering force due to the recoil force of the scattered photons [32]. Optical pulling force is typically achieved by exciting electric dipole moment and magnetic dipole moment or electric quadrupole moment simultaneously so that their interference produces strong forward scattering. We shall see that for a chiral particle, a new mechanism for optical pulling force emerges from the coupling of linear and angular momentums via chirality.

We note that the last two terms of Eq. (10) are proportional to the chirality fluxes $\langle \Psi_e \rangle$ and $\langle \Psi_m \rangle$. This implies that chirality can directly induce a force and it is a new mechanism for manipulating materials with light. This can also be seen in the free energy expression Eq. (6), where the last term of the free energy also depend on chirality. In the paraxial limit, the chirality is proportional to the spin density and the chirality flux induced forces can be interpreted as coupling between the mechanical linear momentum and the spin angular momentum of light. We shall refer such forces as the chirality forces, as it entirely comes from chirality fluxes of the field ($\langle \Psi_e \rangle$ and $\langle \Psi_m \rangle$) and the chirality of the particle ($\alpha_{em}$).

We note that light waves with different chirality will induce opposite chirality forces on the same particle. These chirality forces open up a new opportunity to achieve optical pulling force. When the magnitude of the chirality forces is greater than the summations of the first three terms in Eq. (5), optical pulling force can be achieved.

We shall now verify the above analytical results using the Maxwell stress tensor and the multiple scattering theory [42], which is also known as the generalized Mie Scattering theory for multi-spheres. Such formulation is exact within classical electrodynamics, and subject only to numerical truncation error. Consider the structure shown in Fig. 1(a), which is a collection of metallic spheres whose center is arranged to sit on a left-handed spiral (outlined by a black line). All the structural details are tabulated in the caption of the figure. The incident waves consist of two incoherent plane waves propagating along $+z$ (right circularly polarized, RCP) and $-z$ (left circularly polarized, LCP), with a somewhat arbitrarily chosen $\lambda = 337$ nm. In such a configuration, the first three terms in Eq. (5) are eliminated due to the counter propagation of the two plane waves, but the chirality fluxes for the two plane waves add up, resulting in a non-zero chiral force. We show in Fig. 1(b) the optical forces calculated for different sizes of the spheres that falls in the range of 40 nm$<d<$ 80 nm and the total length of the spiral is set to scale with the size of the spheres. We see that the $z$-component of the force ($F_z$) induced by positive spin angular momentum flux (P-SAMF, $\hat{z} \cdot \langle L_s \rangle > 0$) and negative SAMF ($\hat{z} \cdot \langle \vec{L}_s \rangle < 0$) are equal and opposite to each other and we note in particular that $F_z$ associated with P-SAMF can be in the same or opposite direction to the P-SAMF, implying that the chirality force can be along $+\mathbf{k}$ or $-\mathbf{k}$ depending on the size of the spiral structure. This clearly demonstrates that chirality force can potentially induce a "pulling" force. We note that the parameters here, such as the radius of the spiral, pitch of the spiral, size of the particles etc. are chosen somewhat arbitrarily. Since chirality is a symmetry property, the same physics should be valid in a broad parameter range.

As described above, chirality force can achieve optical pulling force. Fig. 2 shows the optical force for a specific chiral system illuminated by linearly (LP-BB, solid gray line), left circularly (LCP-BB, solid blue line), and right circularly (RCP-BB, solid red line) polarized non-diffractive Bessel beams, which shows that both RCP and LCP Bessel beams can induce a negative force. When the particles and the entire spiral structure are both small, the chiral features cannot be resolved by the incident wave and no optical pulling force is observed. But as the particle and the spiral sizes increase, Fig. 2 shows that circular polarized beams can induce optical forces of opposite signs (see e.g. at sphere diameters of 55 and 70 nm), leading to an optical pulling force for one of the circular polarization. This is in good agreement with our previous predictions, where circularly polarized beams can enhance the optical pulling force. It is important to note that we have achieved optical pulling force for metallic spiral systems in which the chiral object is small, which is very different from previously reported case of non-chiral dielectrics particles [32], where optical pulling force can only be observed in the Mie regime where the dielectric sphere has to be big enough for the electric dipole, magnetic dipole, and electric quadrupole to be simultaneously excited. In order to illustrate the importance of chirality, we also calculate the optical force for planar structures (with six evenly spaced spheres sitting on a circle on the $xy$ plane) illuminated by Bessel beams with different polarization, as shown by dash lines in Fig.2. Clearly, no optical pulling force is observed and there are no difference between RCP-BB and LCP-BB.

This chirality force can be considered as a consequence of coupling between linear momentum and angular momentum. In principles, the reverse operation is also allowed, namely illumination by a linearly polarized beams should induce an optical torque on the chiral particle [19], due to the same coupling effect between linear momentum and angular momentum. This is illustrated by Fig. 3, which demonstrates the existence of the optical torque (about the central axis of the spiral) for the structure shown in Fig. 1a illuminated by a linearly polarized plane wave. The corresponding optical torques for planar structures is also shown, which is strictly zero as there is no chirality.

In conclusion, we have derived an analytical expression for optical forces acting on optically small chiral particles, which has chirality dependent terms that can be completely expressed in terms of the chirality parameter of the electromagnetic wave

$K_{Chirality}$ and the chirality fluxes $\langle \mathbf{\Psi}_e \rangle$ and $\langle \mathbf{\Psi}_m \rangle$. These chirality dependent terms enable a new mechanism to achieve optical pulling force using the coupling between linear momentum of the particle and spin angular momentum of light. Optical pulling force predicted by the analytic theory is explicitly demonstrated in numerical calculations employing multiple scattering method and Maxwell stress tensor.

**Acknowledgement**

This work is supported by Hong Kong RGC grant HKUST2/CRF/11G. KD and JN are also supported by GRF grant 604011.

∗ jacktfng@hkbu.edu.hk

† phchan@ust.hk

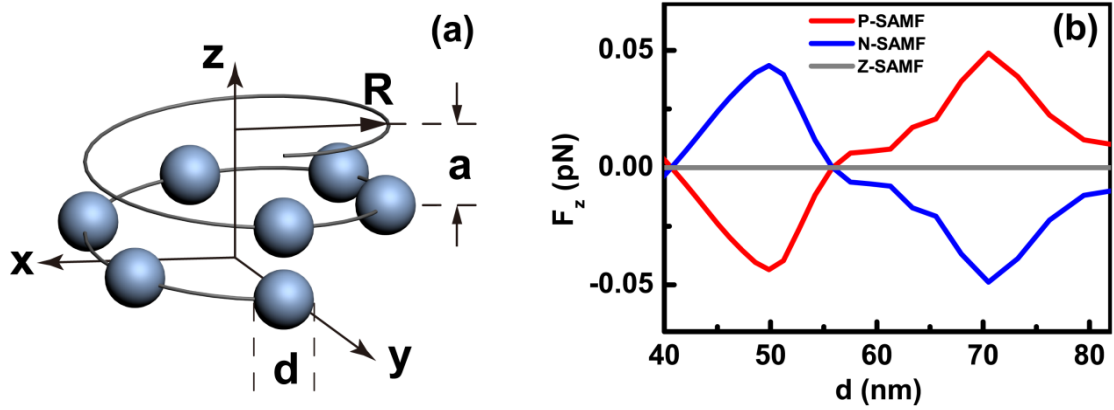

Figure 1. (color online) (a) Schematic picture for the prototypical chiral chain making up of 25 metallic spheres ($\varepsilon_r = -5 + 0.13i$) that are arranged on a spiral marked by the black line and parametrized by $(R\sin 2\pi t, R\cos 2\pi t, at)$, with $0 < t < 4$, and $R = 1.05d$, pitch $a = 1.05 d$ where $d$ is the diameter of the sphere. The spiral's height is $4a$ and each loop has 6 spheres evenly spaced on the spiral. (b) Calculated optical forces on the spiral chains for two counter propagating incoherent plane waves with $\lambda$ = 337 nm propagating in $+z$ and $-z$ direction, each has an intensity of $10^9$ W/m$^2$. Positive spin angular momentum flux (P-SAMF) (with $\langle \Phi_e \rangle \parallel \mathbf{k}$) can induce both positive and negative optical force as shown by the red line. Negative force here means that the beam attracts the object. The blue line shows the force due to a negative spin angular momentum flux (N-SAMF).

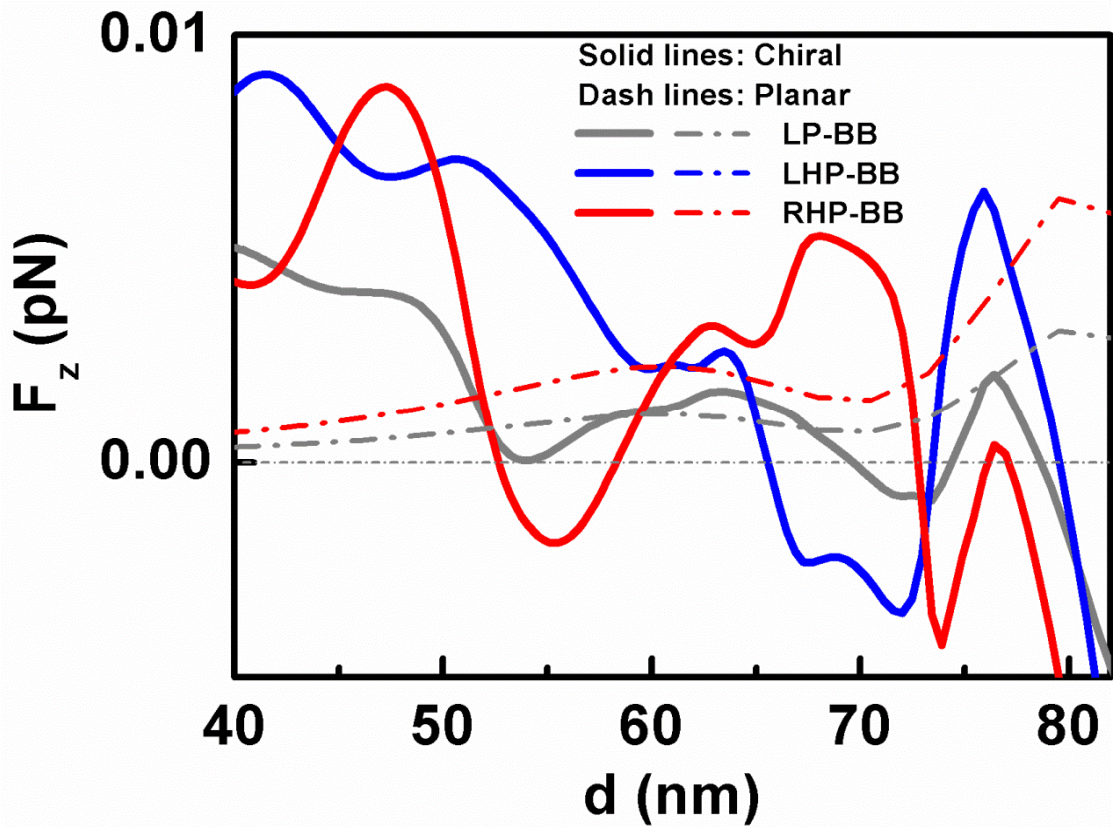

Figure 2. (color online) The *z*-component of the optical forces acting on the chiral chain shown in Fig. 1a, induced by a Bessel beam with $\alpha = 87°$ and $l = 0$ (see supplemental information for the expression of the Bessel beam) with different polarizations. The solid gray, blue, and red lines are for Bessel beams with linear (LP-BB), left circular (LCP-BB ), and right circular polarization (RCP-BB), respectively. The intensity of each beam is $10^9$ W/m$^2$. For comparison, the dashed lines show the optical forces acting on planar structures with the spheres positions projected onto the xy plane. The results show that circularly polarized Bessel beams can induce negative forces on chiral structures (solid lines) while no negative forces can be induced on non-chiral structures (dashed lines).

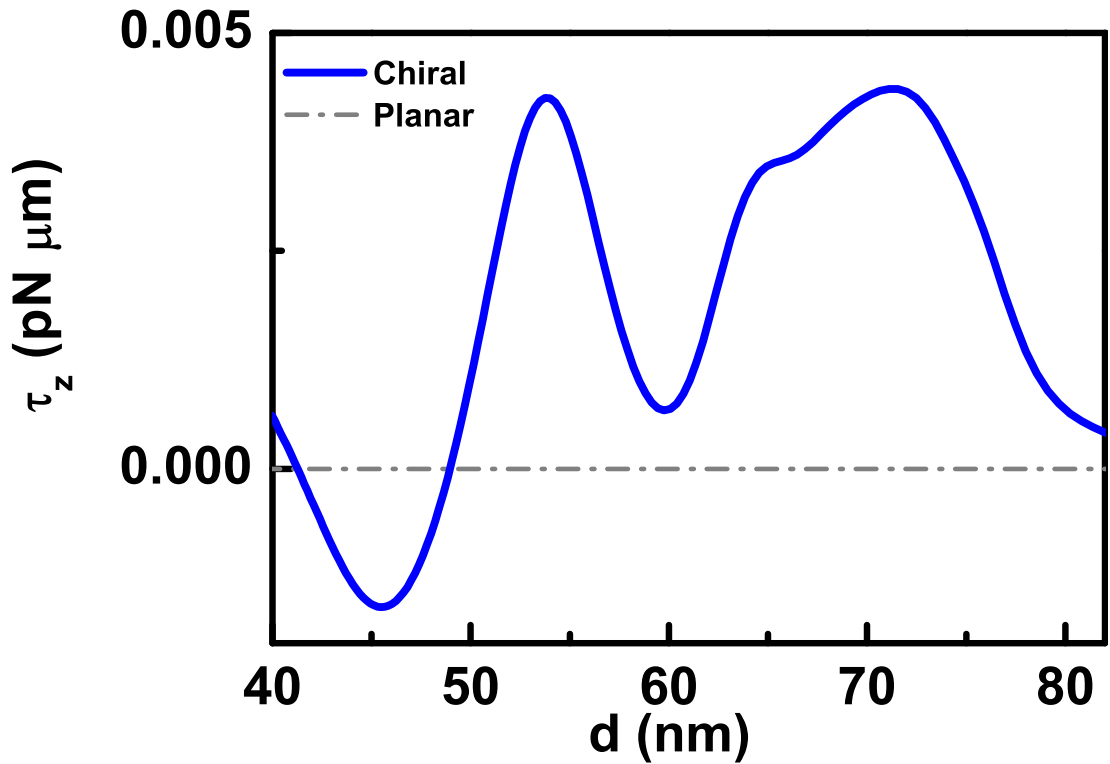

Figure 3. (color online) The *z*-component of the total optical torque about the z-axis for the chiral chain system shown in Fig. 1a, illuminated by an *x*-polarized plane wave (blue line). The intensity of the beam is $10^9$ W/m$^2$. For comparison, the total optical torque acting on a non-chiral planar structures, namely all spheres sitting on *xy*-plane, are shown in the dash lines.

# Supplemental Material:
# Realizing optical pulling force using chirality

Kun Ding,[1,2] Jack Ng,[1,3,*] Lei Zhou,[2] and C. T. Chan[1,†]

[1]Department of Physics and Institute for Advanced Study, The Hong Kong University of Science and Technology, Clear Water Bay, Hong Kong

[2]Department of Physics, Fudan University, Shanghai, China

[3]Department of Physics and Institute of Computational and Theoretical Studies, Hong Kong Baptist University, Kowloon Tong, Kowloon, Hong Kong

## 1. The multipole expansion of optical force acting on a chiral particle

A small chiral particle may be described by an electric dipole moment $\mathbf{p}$ and a magnetic dipole moment $\mathbf{m}$, which are given by [1]

$$\begin{aligned} \mathbf{p} &= \vec{\boldsymbol{\alpha}}^{ee}\mathbf{E} + \vec{\boldsymbol{\alpha}}^{em}\mathbf{B}, \\ \mathbf{m} &= -\vec{\boldsymbol{\alpha}}^{emT}\mathbf{E} + \vec{\boldsymbol{\alpha}}^{mm}\mathbf{B}. \end{aligned} \quad (1)$$

Here $\vec{\boldsymbol{\alpha}}^{ee}$, $\vec{\boldsymbol{\alpha}}^{mm}$, and $\vec{\boldsymbol{\alpha}}^{em}$ are the electric, magnetic, and chiral polarization tensors, $\mathbf{A}^T$ denotes the transpose of $\mathbf{A}$, and $\vec{\boldsymbol{\alpha}}^{me} = -\vec{\boldsymbol{\alpha}}^{emT}$ for a reciprocal medium. The multipole expansion of the time averaged optical force (referred as optical force hereafter) for a particle illuminated by a monochromatic field is given in Ref. 2. Here, we assume that the particle is small compare to the incident wavelength, therefore we keep only the electric and magnetic dipole moments. In this case, the optical force is given by

$$\langle \mathbf{F} \rangle = \mathbf{F_p} + \mathbf{F_m} + \mathbf{F_{p\times m}}, \quad (2)$$

where

$$\mathbf{F_p} = \frac{1}{2}\operatorname{Re}\{(\nabla \mathbf{E}^*)\cdot \mathbf{p}\} \quad (3)$$

is the electric dipole force,

$$\mathbf{F_m} = \frac{1}{2}\operatorname{Re}\{(\nabla \mathbf{B}^*)\cdot \mathbf{m}\}, \quad (4)$$

is the magnetic dipole force, and

$$\mathbf{F}_{\mathbf{p}\times\mathbf{m}} = -\frac{k^4}{12\pi\epsilon_0 c}\text{Re}\{\mathbf{p}\times\mathbf{m}^*\} \tag{5}$$

is the recoil force. We shall re-write Eq. (2) term by term in order to transform it into a compact form.

## 2. Electric dipole terms

The electromagnetic field can be written in Cartesian coordinates:

$$\mathbf{E} = |E_j|\text{e}^{i\Theta_j}\hat{\mathbf{x}}_j, \qquad \mathbf{B} = |B_j|\text{e}^{i\Phi_j}\hat{\mathbf{x}}_j, \tag{6}$$

where $|E_j|$ and $|B_j|$ are the magnitudes of the field's $j$-th Cartesian components, $\Theta_j$ and $\Phi_j$ are the phases of the corresponding fields, and $\hat{\mathbf{x}}_j$ is the $j$-th unit vector in the Cartesian coordinate system. We adopt the Einstein summation convention. Substituting Eq. (6) into Eq. (3), one arrives at

$$\begin{aligned}\mathbf{F}_{\mathbf{p}} &= \tfrac{1}{2}\text{Re}\{(\nabla\mathbf{E}^*)\cdot\mathbf{p}\} = \tfrac{1}{2}\text{Re}\{p_i\nabla E_i^*\} \\ &= \tfrac{1}{2}\text{Re}\{(\alpha_{ij}^{ee}E_j + \alpha_{ik}^{em}B_k)\nabla E_i^*\} \\ &= \tfrac{1}{2}\hat{e}_n\text{Re}\{(\partial_n|E_i|\text{e}^{-i\Theta_i})(\alpha_{ij}^{ee}|E_j|\text{e}^{i\Theta_j} + \alpha_{ik}^{em}|B_k|\text{e}^{i\Phi_k})\} \\ &= \tfrac{1}{2}\hat{e}_n\left\{\begin{array}{l}(\partial_n|E_i|)|E_j|\text{Re}[\alpha_{ij}^{ee}\text{e}^{i(\Theta_j-\Theta_i)}] + (\partial_n|E_i|)|B_k|\text{Re}[\alpha_{ik}^{em}\text{e}^{i(\Phi_k-\Theta_i)}] \\ +|E_i||E_j|(\partial_n\Theta_i)\text{Im}[\alpha_{ij}^{ee}\text{e}^{i(\Theta_j-\Theta_i)}] + |E_i||B_k|(\partial_n\Theta_i)\text{Im}[\alpha_{ik}^{em}\text{e}^{i(\Phi_k-\Theta_i)}]\end{array}\right\}\end{aligned} \tag{7}$$

where $\alpha_{ij}^{ee}$ is the $ij$-component of $\ddot{\boldsymbol{\alpha}}^{ee}$, and similar notations are employed for the other polarizabilities. The first two terms originate from the inhomogeneity of the field amplitude (i.e. $\partial_n|E_i|$), whereas the last two terms originate from the inhomogeneity of the field's phases (i.e. $\partial_n\Theta_i$).

After some manipulations, Eq.(7) can be written in a compact form

$$\mathbf{F}_{\mathbf{p}} = -\nabla\langle U_{\mathbf{p}}\rangle + \tfrac{1}{4}\text{Re}\{\mathbf{T}^{ee}:\ddot{\boldsymbol{\alpha}}^{ee} + \mathbf{T}^{em}:\ddot{\boldsymbol{\alpha}}^{em}\} \tag{8}$$

where

$$\langle U_{\mathbf{p}} \rangle = -\frac{1}{4}\mathrm{Re}\left(\mathbf{E} \cdot \mathbf{p}^*\right), \tag{9}$$

and the auxiliary quantities are given by

$$\begin{aligned}
T^{ee}_{n,ji} &= \left[\left(\partial_n |E_i|\right)|E_j| - \left(\partial_n |E_j|\right)|E_i| - i|E_i||E_j|\partial_n\left(\Theta_i + \Theta_j\right)\right]e^{i(\Theta_j - \Theta_i)}, \\
T^{em}_{n,ki} &= \left[\left(\partial_n |E_i|\right)|B_k| - \left(\partial_n |B_k|\right)|E_i| - i|E_i||B_k|\partial_n\left(\Theta_i + \Phi_k\right)\right]e^{i(\Phi_k - \Theta_i)},
\end{aligned} \tag{10}$$

Here $\mathbf{T}^{ee}$ and $\mathbf{T}^{em}$ are rank 3 tensors that depend only on the field, whereas $\ddot{\alpha}^{ee}$ and $\ddot{\alpha}^{em}$ depend only on the particle, i.e. the contributions from the field and from the particle are separated.

## 3. Magnetic dipole terms

Similar to electric dipole force $\mathbf{F_p}$, the magnetic dipole force $\mathbf{F_m}$ can be obtained by substituting (6) into (4):

$$\begin{aligned}
\mathbf{F_m} &= \frac{1}{2}\mathrm{Re}\left\{(\nabla \mathbf{B}^*) \cdot \mathbf{m}\right\} = \frac{1}{2}\mathrm{Re}\left\{m_i \nabla B_i^*\right\} \\
&= \frac{1}{2}\mathrm{Re}\left\{\left(\alpha_{ij}^{mm} B_j - \alpha_{ik}^{em,\mathrm{T}} E_k\right)\nabla B_i^*\right\} \\
&= \frac{\hat{\mathbf{x}}_n}{2}\mathrm{Re}\left\{\left(\partial_n |B_i| e^{-i\Phi_i}\right)\left(\alpha_{ij}^{mm}|B_j|e^{i\Phi_j} - \alpha_{ik}^{em,\mathrm{T}}|E_k|e^{i\Theta_k}\right)\right\} \\
&= \frac{\hat{\mathbf{x}}_n}{2}\left\{\begin{array}{l}\left(\partial_n |B_i|\right)|B_j|\mathrm{Re}\left[\alpha_{ij}^{mm} e^{i(\Phi_j - \Phi_i)}\right] - \left(\partial_n |B_i|\right)|E_k|\mathrm{Re}\left[\alpha_{ik}^{em,\mathrm{T}} e^{i(\Theta_k - \Phi_i)}\right] \\ + |B_i||B_j|(\partial_n \Phi_i)\mathrm{Im}\left[\alpha_{ij}^{mm} e^{i(\Phi_j - \Phi_i)}\right] - |B_i||E_k|(\partial_n \Phi_i)\mathrm{Im}\left[\alpha_{ik}^{em,\mathrm{T}} e^{i(\Theta_k - \Phi_i)}\right]\end{array}\right\}
\end{aligned} \tag{11}$$

Upon performing similar derivations as in the electric dipole term, we arrive at

$$\mathbf{F_m} = -\nabla\langle U_{\mathbf{m}}\rangle + \frac{1}{4}\mathrm{Re}\left\{\mathbf{T}^{mm}:\ddot{\alpha}^{mm} + \mathbf{T}^{me}:\ddot{\alpha}^{em,\mathrm{T}}\right\} \tag{12}$$

where

$$\langle U_{\mathbf{m}} \rangle = -\frac{1}{4}\mathrm{Re}\left(\mathbf{B} \cdot \mathbf{m}^*\right), \tag{13}$$

and the auxiliary quantities are given by

$$\begin{aligned}
T^{mm}_{n,ji} &= \left[\left(\partial_n |B_i|\right)|B_j| - \left(\partial_n |B_j|\right)|B_i| - i|B_i||B_j|\partial_n\left(\Phi_i + \Phi_j\right)\right]e^{i(\Phi_j - \Phi_i)}, \\
T^{me}_{n,ki} &= \left[\left(\partial_n |E_k|\right)|B_i| - \left(\partial_n |B_i|\right)|E_k| + i|B_i||E_k|\partial_n\left(\Phi_i + \Theta_k\right)\right]e^{i(\Theta_k - \Phi_i)},
\end{aligned} \tag{14}$$

The expression Eq. (12) is similar to its electric counterpart Eq. (8).

## 4. The $\mathbf{p} \times \mathbf{m}$ term

Using Eq. (6), the recoil force term Eq. (5) can be written as

$$\begin{aligned}\mathbf{F}_{\mathbf{p}\times\mathbf{m}} &= -\frac{k^4}{12\pi\epsilon_0 c}\operatorname{Re}\{\mathbf{p}\times\mathbf{m}^*\} \\ &= -\frac{k^4}{12\pi\epsilon_0 c}\operatorname{Re}\{(\vec{\boldsymbol{\alpha}}^{ee}\mathbf{E}+\vec{\boldsymbol{\alpha}}^{em}\mathbf{B})\times(-\vec{\boldsymbol{\alpha}}^{em\dagger}\mathbf{E}^*+\vec{\boldsymbol{\alpha}}^{mm*}\mathbf{B}^*)\}\end{aligned} \quad (15)$$

where $\mathbf{A}^\dagger$ denotes the Hermitian conjugate of $\mathbf{A}$. Using the mathematical identity:

$$\begin{aligned}(\vec{\mathbf{A}}\cdot\vec{\boldsymbol{\mu}})\times(\vec{\mathbf{B}}\cdot\vec{\mathbf{v}}) &= \left[(\vec{\mathbf{A}}\cdot\vec{\boldsymbol{\mu}})\times\vec{\mathbf{B}}\right]\cdot\vec{\mathbf{v}} = -\left[\vec{\mathbf{B}}^T\times(\vec{\mathbf{A}}\cdot\vec{\boldsymbol{\mu}})\right]^T\cdot\vec{\mathbf{v}} \\ &= -\left[(\vec{\mathbf{B}}^T\times\vec{\mathbf{A}})\cdot\vec{\boldsymbol{\mu}}\right]^T\cdot\vec{\mathbf{v}} = -\vec{\boldsymbol{\mu}}^T\cdot(\vec{\mathbf{B}}^T\times\vec{\mathbf{A}})^T\cdot\vec{\mathbf{v}}\end{aligned} \quad (16)$$

one can transform Eq. (15) into

$$\begin{aligned}\mathbf{F}_{\mathbf{p}\times\mathbf{m}} &= -\frac{k^4}{12\pi\epsilon_0 c}\operatorname{Re}\{(\vec{\boldsymbol{\alpha}}^{ee}\mathbf{E}+\vec{\boldsymbol{\alpha}}^{em}\mathbf{B})\times(-\vec{\boldsymbol{\alpha}}^{em\dagger}\mathbf{E}^*+\vec{\boldsymbol{\alpha}}^{mm*}\mathbf{B}^*)\} \\ &= -\frac{k^4}{12\pi\epsilon_0 c}\operatorname{Re}\left\{\begin{array}{l}(\vec{\boldsymbol{\alpha}}^{ee}\mathbf{E})\times(-\vec{\boldsymbol{\alpha}}^{em\dagger}\mathbf{E}^*)+(\vec{\boldsymbol{\alpha}}^{ee}\mathbf{E})\times(\vec{\boldsymbol{\alpha}}^{mm*}\mathbf{B}^*) \\ +(\vec{\boldsymbol{\alpha}}^{em}\mathbf{B})\times(-\vec{\boldsymbol{\alpha}}^{em\dagger}\mathbf{E}^*)+(\vec{\boldsymbol{\alpha}}^{em}\mathbf{B})\times(\vec{\boldsymbol{\alpha}}^{mm*}\mathbf{B}^*)\end{array}\right\} \\ &= -\frac{k^4}{12\pi\epsilon_0 c}\operatorname{Re}\left\{\begin{array}{l}\mathbf{E}\cdot(\vec{\boldsymbol{\alpha}}^{em*}\times\vec{\boldsymbol{\alpha}}^{ee})^T\cdot\mathbf{E}^* - \mathbf{E}\cdot(\vec{\boldsymbol{\alpha}}^{mm?}\times\vec{\boldsymbol{\alpha}}^{ee})^T\cdot\mathbf{B} \\ +\mathbf{B}\cdot(\vec{\boldsymbol{\alpha}}^{em*}\times\vec{\boldsymbol{\alpha}}^{em})^T\cdot\mathbf{E}^* - \mathbf{B}\cdot(\vec{\boldsymbol{\alpha}}^{mm?}\times\vec{\boldsymbol{\alpha}}^{em})^T\cdot\mathbf{B}^*\end{array}\right\}\end{aligned} \quad (17)$$

Introducing two more auxiliary quantities

$$\begin{aligned}\tilde{\mathbf{L}} &= \mathbf{E}\cdot(\vec{\boldsymbol{\alpha}}^{em*}\times\vec{\boldsymbol{\alpha}}^{ee})^T\cdot\mathbf{E}^* - \mathbf{B}\cdot(\vec{\boldsymbol{\alpha}}^{mm?}\times\vec{\boldsymbol{\alpha}}^{em})^T\cdot\mathbf{B} , \\ \tilde{\mathbf{S}} &= \mathbf{B}\cdot(\vec{\boldsymbol{\alpha}}^{em*}\times\vec{\boldsymbol{\alpha}}^{em})^T\cdot\mathbf{E}^* - \mathbf{E}\cdot(\vec{\boldsymbol{\alpha}}^{em?}\times\vec{\boldsymbol{\alpha}}^{ee})^T\cdot\mathbf{B} ,\end{aligned} \quad (18)$$

Eq. (17) then becomes

$$\mathbf{F}_{\mathbf{p}\times\mathbf{m}} = -\frac{k^4}{12\pi\epsilon_0 c}\operatorname{Re}\{\tilde{\mathbf{L}}+\tilde{\mathbf{S}}\} \quad (19)$$

## 5. Grouping all terms together

Summing up (8), (12), and (19), the total optical force is given by

$$\langle \mathbf{F} \rangle = -\nabla \langle U_f \rangle + \frac{1}{4} \text{Re} \left\{ \begin{array}{l} \mathbf{T}^{ee} : \ddot{\boldsymbol{\alpha}}^{ee} + \mathbf{T}^{em} : \ddot{\boldsymbol{\alpha}}^{em} \\ + \mathbf{T}^{mm} : \ddot{\boldsymbol{\alpha}}^{mm} + \mathbf{T}^{me} : \ddot{\boldsymbol{\alpha}}^{emT} \end{array} \right\} - \frac{k^4}{12\pi\epsilon_0 c} \text{Re}\{\tilde{\mathbf{L}} + \tilde{\mathbf{S}}\}, \quad (20)$$

$$\langle U_f \rangle = \langle U_\mathbf{p} \rangle + \langle U_\mathbf{m} \rangle = -\frac{1}{4}\text{Re}(\mathbf{E} \cdot \mathbf{p}^*) - \frac{1}{4}\text{Re}(\mathbf{B} \cdot \mathbf{m}^*) \quad (21)$$

This is just Eq. (5) in the main text. In order to grasp the physics, in the next session, we consider a special case of Eq. (20), namely a small isotropic chiral particle.

## 6 Isotropic chiral particles

To understand the essence of chirality induced optical force, we consider a particle with isotropic polarizabilities: $(\ddot{\boldsymbol{\alpha}}^{ee})_{ij} = \alpha_{ee}\delta_{ij}$, $(\ddot{\boldsymbol{\alpha}}^{em})_{ij} = \alpha_{em}\delta_{ij}$, , and $(\ddot{\boldsymbol{\alpha}}^{mm})_{ij} = \alpha_{mm}\delta_{ij}$. With these, Eq. (20) is simplified to

$$\langle \mathbf{F} \rangle = -\nabla \langle U_f \rangle + (C_{ext} + C_{recoil})c^{-1}\langle \mathbf{S} \rangle$$
$$+ \nabla \times \left[ C^\mathbf{p}_{ext} c \langle \mathbf{L}^\mathbf{p}_s \rangle + C^\mathbf{m}_{ext} c \langle \mathbf{L}^\mathbf{m}_s \rangle + \mu_0 (\text{Re}\,\alpha_{em})\langle \mathbf{S} \rangle \right]$$
$$+ \left[ 2\omega^2 \mu_0 (\text{Re}\,\alpha_{em}) - \frac{k^5}{3\pi\epsilon_0} \frac{\text{Im}(\alpha_{ee}\alpha^*_{em})}{\epsilon_0} \right] \langle \mathbf{L}^\mathbf{p}_s \rangle \quad (22)$$
$$+ \left[ 2\omega^2 \mu_0 (\text{Re}\,\alpha_{em}) - \frac{k^5 \mu_0}{3\pi\epsilon_0} \text{Im}(\alpha_{mm}\alpha^*_{em}) \right] \langle \mathbf{L}^\mathbf{m}_s \rangle$$

where

$$\langle U_f \rangle = -\frac{1}{4}\text{Re}(\alpha_{ee})|\mathbf{E}|^2 - \frac{1}{4}\text{Re}(\alpha_{mm})|\mathbf{B}|^2 + \frac{1}{2}\text{Im}(\alpha_{em})\text{Im}(\mathbf{B}\cdot\mathbf{E}^*) \quad (23)$$

is the free energy,

$$C_{ext} = C^\mathbf{p}_{ext} + C^\mathbf{m}_{ext} \quad (24)$$

is the extinction cross section,

$$C^\mathbf{p}_{ext} = \frac{k\,\text{Im}(\alpha_{ee})}{\epsilon_0}, \quad (25)$$

is the extinction cross section through the electric dipole channel,

$$C^\mathbf{m}_{ext} = k\mu_0 \text{Im}(\alpha_{mm}) \quad (26)$$

is the extinction cross section through the magnetic dipole channel,

$$C_{recoil} = -\frac{k^4 \mu_0 \text{Re}(\alpha_{ee}\alpha^*_{mm})}{6\pi\epsilon_0} - \frac{k^4 \mu_0}{6\pi\epsilon_0}\text{Re}(\alpha_{em}\alpha^*_{em}) \quad (27)$$

is directly related to the recoil force,

$$\langle \mathbf{L}_S^m \rangle = \frac{\mu_0}{4\omega i} \mathbf{H} \times \mathbf{H}^*, \tag{28}$$

and

$$\langle \mathbf{L}_s^p \rangle = \frac{\epsilon_0}{4\omega i} \mathbf{E} \times \mathbf{E}^*. \tag{29}$$

We note that the chirality flux

$$\mathbf{F} = \mu_0 \omega \left[ \varepsilon_0 \mathbf{E}^* \times \mathbf{E} + \mu_0 \mathbf{H}^* \times \mathbf{H} \right] = -4\omega^2 \mu_0 \operatorname{Re}\{\langle \mathbf{L}_s^p \rangle + \langle \mathbf{L}_s^m \rangle\} \tag{30}$$

is directly related to Eq. (28) and Eq. (29), indicating a close physical relation between chirality and optical forces.

## 7 Paraxial Approximation

For paraxial beams or plane waves, $\mathbf{k} // \mathbf{E} \times \mathbf{E}^*$ where $\mathbf{k}$ is the propagation direction of the incident wave, one has

$$\langle \mathbf{L}_s^m \rangle = \frac{\mu_0}{4\omega i} \mathbf{H} \times \mathbf{H}^* = \frac{\epsilon_0}{4\omega i} \hat{\mathbf{k}}[\hat{\mathbf{k}} \cdot (\mathbf{E} \times \mathbf{E}^*)] = \langle \mathbf{L}_S^p \rangle \equiv \langle \mathbf{L}_s \rangle. \tag{31}$$

Using Eq. (31) and assuming $|\operatorname{Re}\alpha_{em}| \ll |\operatorname{Im}\alpha_{em}|$, Eq. (22) can be expressed in an even more compact form,

$$\begin{aligned}\langle \mathbf{F} \rangle = &-\nabla \langle U_f \rangle + (C_{ext} + C_{recoil}) c^{-1} \langle \mathbf{S} \rangle + C_{ext} c \nabla \times \langle \mathbf{L}_s \rangle \\ &- \frac{k^5}{3\pi\epsilon_0^2} \operatorname{Im}[\alpha_{em}^*(\alpha_{ee} + \frac{\alpha_{mm}}{c^2})] \langle \mathbf{L}_s \rangle\end{aligned}. \tag{32}$$

This is just Eq. (15) in the main text.

## 8 Bessel Beam

In this paper, to eliminate axial gradient force along the $z$-direction, we consider Bessel beams of the form [3]

$$E_x(\rho,\phi,z) = \frac{1}{2} E_0(\alpha_0) e^{ik_z z} e^{il\phi} \begin{cases} P_\perp i^{2-l} J_{2-l}(k_\perp \rho) e^{-i2\phi} (-1 \pm 1) \\ + P_\perp i^{2+l} J_{2+l}(k_\perp \rho) e^{i2\phi} (-1 \mp 1) \\ + i^{-l} J_{-l}(k_\perp \rho) + i^l J_l(k_\perp \rho) \end{cases} \tag{33}$$

$$E_y(\rho,\phi,z) = \frac{1}{2} E_0(\alpha_0) e^{ik_z z} e^{il\phi} \begin{Bmatrix} P_\perp i^{3-l} J_{2-l}(k_\perp \rho) e^{-i2\phi} (-1\pm 1) \\ + P_\perp i^{3+l} J_{2+l}(k_\perp \rho) e^{i2\phi} (1\pm 1) \\ + i^{1-l} J_{-l}(k_\perp \rho) + i^{1+l} J_l(k_\perp \rho) \end{Bmatrix} \quad (34)$$

$$E_z(\rho,\phi,z) = E_0(\alpha_0) e^{ik_z z} e^{il\phi} P_\parallel \begin{Bmatrix} i^{1-l} J_{1-l}(k_\perp \rho) e^{-i\phi} (-1\pm 1) \\ + i^{1+l} J_{1+l}(k_\perp \rho) e^{i\phi} (-1\mp 1) \end{Bmatrix} \quad (35)$$

where $E_0$ is the on-axis electric field, $J_n$ is the Bessel function of $n$-th order, $(\rho,\phi,z)$ are cylindrical coordinates, $\pm(\mp)$ means left (right) handed circular polarization and

$$\begin{aligned} k_\perp &= k\sin\alpha_0 & k_z &= k\cos\alpha_0 \\ P_\perp &= \frac{1-\cos\alpha_0}{1+\cos\alpha_0} & P_\parallel &= \frac{\sin\alpha_0}{1+\cos\alpha_0} \end{aligned} \quad (36)$$

In the calculation of Fig.2, the parameters we used are $\alpha_0 = 87°$ and $l = 0$.